\documentclass{rnaastex}

\begin{document}

\title{LP 543-25: a rare low-mass runaway disk star}

\correspondingauthor{Ra\'ul~de~la~Fuente~Marcos}
\email{rauldelafuentemarcos@ucm.es}

\author[0000-0002-5319-5716]{Ra\'ul~de~la~Fuente~Marcos}
\affiliation{AEGORA Research Group,
             Facultad de Ciencias Matem\'aticas,
             Universidad Complutense de Madrid, 
             Ciudad Universitaria, E-28040 Madrid, Spain}

\author[0000-0003-3894-8609]{Carlos~de~la~Fuente~Marcos}
\affiliation{Universidad Complutense de Madrid,
             Ciudad Universitaria, E-28040 Madrid, Spain}


\section{} 

LP~543-25 or PSS~544-7 is a high proper-motion star, which was discovered by \citet{1991AJ....102..395H} during a survey with the 2.1~m 
telescope at Kitt Peak National Observatory. LP~543-25 attracted some attention due to its high tangential velocity \citep{2005NewA...10..551D,
2005AJ....129.1483L}; \citet{2005NewA...10..551D} argued that given its location and kinematic signature, LP~543-25 could be a candidate 
cannonball star ejected by a star cluster. However, the input data used in the original analysis by \citet{2005NewA...10..551D} were 
rather uncertain and so were the results: a tangential velocity of $399\pm127$~km~s$^{-1}$ and a $W$-component $>350$~km~s$^{-1}$. Here, 
we revisit the issue of the kinematics of this interesting star using {\it Gaia} DR2.

{\it Gaia} DR2 \citep{2016A&A...595A...1G,2018arXiv180409365G} provides the necessary, high-quality data ---right ascension and declination, 
stellar parallax, radial velocity, proper motions in right ascension and declination, and their respective standard errors--- to investigate 
the kinematics of LP~543-25, which is source {\it Gaia} DR2 3137866242753594624. Such an input dataset can be used to compute Galactic space 
velocities and their uncertainties following the procedure outlined by \citet{1987AJ.....93..864J}. Here, we consider values of the Solar 
motion from \citet{2010MNRAS.403.1829S} and use a right-handed coordinate system for $U$, $V$, and $W$; axes are positive in the directions 
of the Galactic Center, Galactic rotation, and the North Galactic Pole.  

\begin{figure}[!ht]
\begin{center}
\includegraphics[scale=0.37,angle=0]{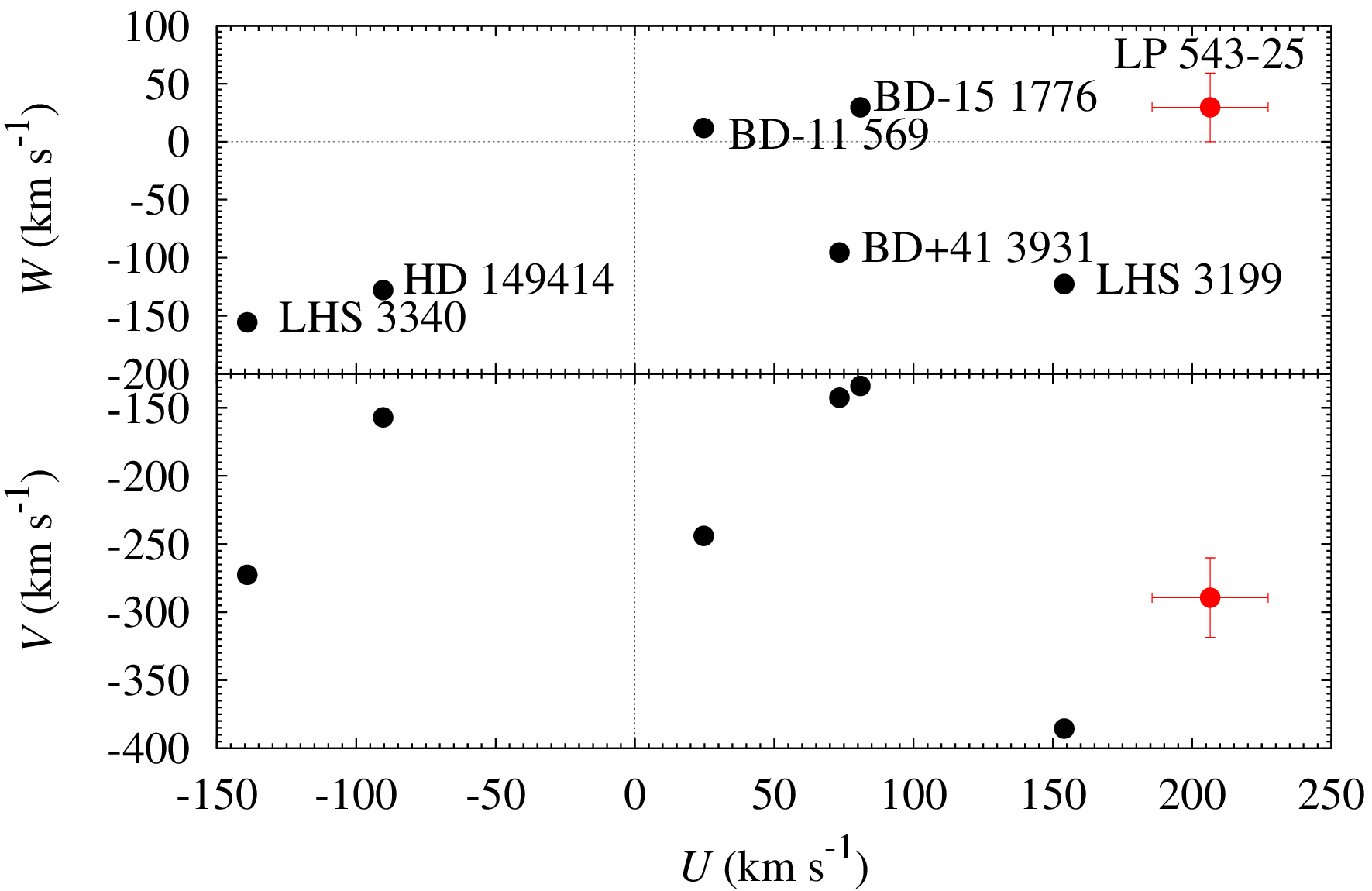}
\includegraphics[scale=0.37,angle=0]{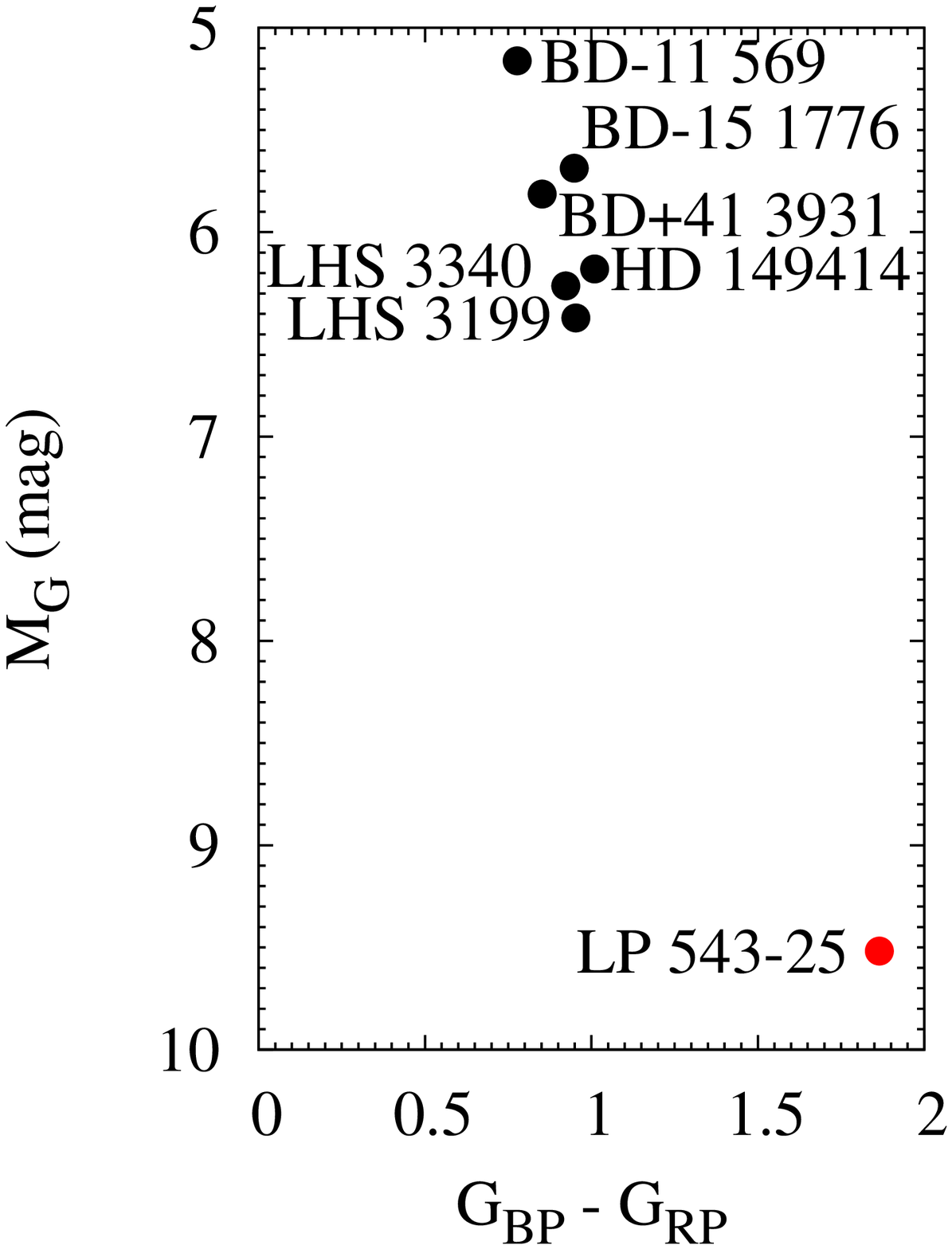}
\caption{Heliocentric Galactic velocity components of LP~543-25 (in red) and six runaway stars of spectral type G (in black), left-hand side
         panels. Hertzsprung-Russell diagram of the same stars, similar to fig. 5 in \citet{2018arXiv180409378G}, right-hand side panel. 
         The input data used to plot this figure are from {\it Gaia} DR2. 
\label{fig:1}}
\end{center}
\end{figure}

{\it Gaia} DR2 does not list a value for the radial velocity of LP~543-25, therefore we will assume that its radial motion is negligible in
comparison with its tangential one; its distance is $458_{-42}^{+51}$~pc. Computed as described above, the Heliocentric Galactic velocity 
components are $U = 206\pm21$~km~s$^{-1}$, $V = -289\pm29$~km~s$^{-1}$, and $W = 30\pm30$~km~s$^{-1}$ (Figure~\ref{fig:1}, left-hand side 
panels); the corresponding Galactocentric Galactic velocity components are $U = -195\pm21$~km~s$^{-1}$, $V = -34\pm29$~km~s$^{-1}$, and 
$W = 37\pm30$~km~s$^{-1}$. The star is therefore moving in the Galactic plane and away from the Galactic Center at a rate of nearly 
200~km~s$^{-1}$, i.e. it is bound to the Galaxy. With such kinematic properties it is not unreasonable to speculate that LP~543-25 may come 
from any of the multiple star clusters that inhabit the inner regions of the Milky Way. \citet{1990AJ.....99..608L} showed numerically that 
star clusters can produce dynamically ejected runaway stars with a maximum velocity close to or above 200~km~s$^{-1}$. A preliminary 
analysis of the photometric data of this star provided by {\it Gaia} DR2 (Figure~\ref{fig:1}, right-hand side panel) confirms the 
conclusions in \citet{2005NewA...10..551D}: LP~543-25 could be a late (Population~I) K dwarf/early M dwarf or a Population~II 
dwarf/subdwarf.

LP~543-25 appears to be a member of an elusive class of stars, the low-mass runaway stars (see e.g. \citealt{2005ApJ...627L..61P,
2015MNRAS.447.2046H}). \citet{2005ApJ...627L..61P} discussed three low-mass runaway stars from the Orion Nebula, but {\it Gaia} DR2 does not 
provide data on any of these objects. The catalog of young runaway stars compiled by \citet{2011MNRAS.410..190T} includes six stars 
---BD-11~569, BD-15~1776, BD+41~3931, LHS~3199, LHS~3340, and HD~149414--- with probable masses under 1~$M_{\odot}$ and tangential velocity 
above 100~km~s$^{-1}$. These six stars have data in {\it Gaia} DR2 and they are plotted in Figure~\ref{fig:1}. Using these well-studied 
runaway stars, it is possible to confirm that LP~543-25 is indeed a low-mass runaway star, perhaps one of the closest and less massive 
runaway stars identified so far. Although certainly promising, these results must be considered with some caution, spectroscopic information, 
including radial velocity determination, is required in order to constrain better the overall properties of this interesting star.

\acknowledgments

This research has made use of the SIMBAD database and the VizieR catalogue access tool, operated at CDS, Strasbourg, France.
This work was partially supported by the Spanish MINECO under grant ESP2015-68908-R. In preparation of this Note, we made use of the NASA 
Astrophysics Data System. This work has made use of data from the European Space Agency (ESA) mission {\it Gaia} 
(\url{https://www.cosmos.esa.int/gaia}), processed by the {\it Gaia} Data Processing and Analysis Consortium (DPAC, 
\url{https://www.cosmos.esa.int/web/gaia/dpac/consortium}). Funding for the DPAC has been provided by national institutions, in particular 
the institutions participating in the {\it Gaia} Multilateral Agreement.

\end{document}